\documentstyle[preprint,epsf,floats,aps,here]{revtex}

\begin{document}

\draft
 
\title{Classical Kolmogorov scaling is inconsistent with local 
    coupling}
\author{C. Uhlig and J. Eggers}
\address{Fachbereich Physik\\ 
  Universit\"at -- Gesamthochschule -- Essen\\D--45117 Essen, Germany}
\date{\today}

\maketitle
 
\begin{abstract}
We consider cascade models of turbulence which are obtained 
by restricting the Navier-Stokes equation to local interactions.
By combining the results of the  method of extended self-similarity and a 
novel subgrid model, we investigate the inertial range fluctuations 
of the cascade. Significant corrections to the classical scaling
exponents are found. The dynamics of our local Navier-Stokes models
is described accurately by a simple set of Langevin equations 
proposed earlier as a model of turbulence [Phys. Rev. E {\bf 50}, 
285 (1994)]. This allows for a prediction of the intermittency exponents 
without adjustable parameters. Excellent agreement with
numerical simulations is found. 
\end{abstract}

\pacs{PACS-numbers: 47.27.Jv, 02.50.Ey, 47.27.Eq, 47.11.+j}



\section{Introduction}
\label{sec:intro}

Much of our intuitive understanding of turbulence is based on the
concept of interactions which are local in k-space. Physically, 
it is based on the notion that most of the distortion of a turbulence 
element or eddy can only come from eddies of comparable size. 
Turbulent features which are much larger only uniformly translate 
smaller eddies, which does not contribute to the energy transfer.
This immediately leads to the idea of a chain of turbulence elements,
through which energy is transported to the energy dissipating 
scales. Accepting such a cascade structure of the turbulent velocity 
field, it is natural to assume that the statistical average 
of velocity differences ${\bf \delta v(r) = v(x+r) - v(x)}$ over a distance 
$|{\bf r}|$ follows scaling laws
\begin{equation}
  \label{vscaling}
     \left<|{\bf v(x+r) - v(x)}|^p\right> \sim r^{\zeta_p}
\end{equation}
in the limit of high Reynolds numbers. By taking velocity {\it differences}
over a distance $r$, one probes objects of corresponding size. 

In addition to this assumption of self-similarity, Kolmogorov 
\cite{kolmogorov41} also made the seemingly intuitive assumption that 
the local statistics of the velocity field should be {\it independent} 
of large-scale flow features, from which it is widely separated in scale. 
Because the turbulent state is maintained by a mean energy flux 
$\epsilon$, the only local scales available are the length $r$ and
$\epsilon$ itself, which leads to the estimate $\delta v \sim 
(\epsilon r)^{1/3}$ or 
\begin{equation}
  \label{2/3}
    \zeta^{(class)}_p = p/3 \quad.
\end{equation}
At the same time, one obtains an estimate for the Kolmogorov length 
\begin{equation}
  \label{eta}
    \eta = (\nu^3/\epsilon)^{1/4}
\end{equation}
where viscosity is important. However, it was only appreciated later
\cite{landau59} that in turbulence long-range correlations
always exist in spite of local coupling. Namely, large-scale fluctuations 
of the velocity field will result in a fluctuating energy transfer, 
which drives smaller scales. As a result, the statistics of the small-scale 
velocity fluctuations will be influenced by the energy transfer 
and fluctuations on widely separated scales are correlated, 
violating the fundamental assumption implicit in (\ref{2/3}) and
(\ref{eta}). 

Indeed, Kolmogorov \cite{kolmogorov62} and Obukhov \cite{obukhov62} 
later proposed the existence of 
corrections to the scaling exponents (\ref{2/3}), 
\begin{equation}
  \label{correct}
     \zeta_p = p/3 + \delta\zeta_p \;, \delta\zeta_p \ne 0
\end{equation}
which were subsequently confirmed experimentally 
\cite{anselmet84,benzi93a,benzi93b,herweijer95}. 
On one hand, careful laboratory 
experiments have been performed at ever higher Reynolds numbers
\cite{anselmet84,castaing90}. On the other hand, a new method of 
data analysis \cite{benzi93a,benzi93b} has been successful
in eliminating part of the effects of viscosity. In particular, 
for the highest moments up to $p = 18$ significant corrections to 
classical scaling were found, a currently accepted value for the so-called 
intermittency parameter $\mu$ being 
\cite{anselmet84}
\begin{equation}
  \label{mu}
     \mu = -\delta\zeta_6 = 0.2 \quad,
\end{equation}
which is a 10 \% correction. The existence of corrections like 
(\ref{mu}) implies that on small scales large fluctuations are much more 
likely to occur than predicted by classical theory. 

This ``intermittent'' behavior is thus most noticeable in derivatives 
of the velocity field such as the local rate of energy dissipation 
\[
\epsilon({\bf x},t) = -\frac{\nu}{2}\left(\partial u_i/\partial x_k 
             + \partial u_k/\partial x_i\right)^2 \quad.
\]
Much of the research in turbulence has been devoted to the study
of the spatial structure of $\epsilon({\bf x},t)$ 
\cite{kolmogorov62,nelkin89}, but which will not be considered here. 
The statistical average of this quantity is what we simply called 
$\epsilon$ before. Owing to energy conservation, it must be equal 
to the mean energy transfer.

The local coupling structure of turbulence has inspired the study 
of so-called shell models, where each octave in wavenumber is 
represented by a {\it constant} number of modes, which are 
only locally coupled. This allows to focus on the implications of 
local coupling for intermittent fluctuations, disregarding 
effects of convection and mixing. The mode representation 
of a single shell serves as a simple model for the ``coherent
structures'' a turbulent velocity field is composed of, and which 
to date have only been poorly characterized, both experimentally and
theoretically. 

We caution that this leaves out two important 
aspects of turbulence, both of which have recently been proposed
to lie at the heart of intermittent behavior. First, we have assumed
that coherent structure are simple in the sense that they only possess
a single scale. But experimental \cite{douady91}, numerical
\cite{kida92}, and theoretical \cite{moffatt94} evidence points to 
the importance of long and skinny threads of vorticity. Although 
their real significance to turbulence is not without dispute
\cite{jimenez93}, they have led to quantitative predictions of
intermittency exponents \cite{she94}, in excellent agreement 
with experiment. Second, nonlocal interactions are disregarded in 
shell models. On the other hand it has recently been proposed 
\cite{lvov95} in the framework of perturbation theory that non-local
interactions can in fact be re-summed to yield correction exponents.

Given the complexity of the turbulence problem, this makes it all 
the more interesting to carefully assess the possibilities for
intermittent fluctuations in the case where both effects are 
eliminated. Thus the aim of this paper is to combine previous 
numerical \cite{eggers91a,grossmann94a} and analytical 
\cite{eggers92,eggers94} efforts to gain insight into intermittent 
fluctuations in models with local coupling alone. We focus on a
particular class of shell models, introduced in \cite{eggers91a}, 
which establishes a direct connection with the Navier-Stokes equation. 
To this end the Navier-Stokes equation is projected on a 
self-similar selection of Fourier modes, which enforce local 
coupling. We will adopt the name {\bf RE}duced {\bf W}ave vector
set {\bf A}pproximation (REWA) here \cite{grossmann94a}. 

In the present paper, most of our effort is devoted to separating 
inertial range fluctuations from other effects relating to the 
fact that scale invariance is broken either by an external length scale
$L$ or by the Kolmogorov length $\eta$. To this end we simulate 
extremely long cascades, covering up to 5 decades in scale. To 
eliminate viscous effects as much as possible, we use the method of extended 
self-similarity (ESS) \cite{benzi93a,benzi93b}.
In addition, to make sure our results are independent of the method of 
data analysis, we develop a new subgrid model using {\it fluctuating}
eddy viscosities \cite{domaradzki95}. The results are in excellent agreement 
with those found from ESS. Finally, and perhaps most importantly, 
we use an analytical calculation \cite{eggers94} to {\it predict}
the correction exponent without adjustable parameters. Again, we find the 
same values within error bars. We show that the stochastic model 
introduced in \cite{eggers92} gives an excellent description of
a REWA cascade, both in an equilibrium and non-equilibrium state. 
We compare the equilibrium properties to adjust a single free parameter
in the stochastic model. 

Thus we reach two objectives: First, we gain analytical insight into 
the origin of intermittent fluctuations in a local cascade. Second, we 
demonstrate for a simple example how equilibrium information about
the interactions of Fourier modes can be used to compute intermittency
exponents. 

Unlike shell models with only one complex mode per shell 
\cite{obukhov71,gledzer73,yamada87} exponent 
corrections $\delta\zeta_p$ for REWA cascades are quite small
\cite{eggers91a,grossmann94a}. Careful studies of 
inertial range fluctuations have found them to be significant 
\cite{grossmann94a}, but their numerical value is only about 1/10
of the numbers found experimentally \cite{anselmet84}. This has
lead to the idea that perhaps the experimentally observed exponent 
corrections are not genuine inertial range properties, but result from
extrapolations of (\ref{vscaling}) into regimes where stirring or
viscous effects are important \cite{eggers91a,grossmann94a,grossmann93}. 

We will not follow up on this idea here. However, we note that 
the REWA cascade misses an essential feature of three-dimensional 
turbulence and thus can hardly be expected to yield {\it quantitative}
predictions. Namely, in real turbulence, the 
number of modes within a shell proliferates like $r^{-2}$ as one
goes to smaller scales, while in cascade models this number is 
constant. This can be remedied by allowing a particular shell to branch
out into eight sub-shells, which represent eddies of half the original 
size. It has been argued \cite{eggers91b,eggers92} that the 
competition between eddies of the same size is responsible for the much larger 
growth of fluctuations observed in three-dimensional turbulence. 
The difficulty with this approach is that one also has to take 
{\it convection} of spatially localized structures into account. 
Also, it is not obvious how to disentangle interactions which are local 
in $k$-space from those local in real space, as to systematically reduce 
the coupling of the Navier-Stokes equation to a {\it tree} of 
turbulence elements. Recently developed wavelet methods are a promising step
\cite{farge92}, but they have so far been used only for data analysis
\cite{meneveau91}.

The paper is organized as follows: In the next section we introduce 
both the mode-reduced approximations of the Navier-Stokes equation 
and the corresponding Langevin models. The inertial range properties
of the latter have only one adjustable parameter, as explained in the 
third section. This parameter is determined for a given selection 
of Fourier modes by considering the equilibrium fluctuations of both
models. We are thus able to predict intermittency exponents of the 
turbulent state without adjustable parameters. The result is
compared with numerical simulations of the REWA cascades in the fourth
section. Exponents are determined by carefully examining various 
sources of error, and are in excellent agreement with the theoretical 
prediction. In the fifth section, we investigate temporal correlations in 
REWA models to inquire further in the origin of intermittency in 
models with local coupling. This also sheds light on the reasons for the 
success of the simple Langevin model used by us.



\section{Two cascade models}
\label{sec:model}
\subsection{Reduced wave vector set approximation}
The REWA model \cite
{eggers91a,grossmann94a} is based on the full Fourier--transformed
Navier--Stokes equation within a volume of periodicity $(2\pi L)^{3}$. In order
to restrict the excited Fourier--modes of the turbulent velocity field
to a numerically tractable number, the Navier--Stokes equation is projected
onto a self--similar set of wave vectors ${\cal K}=\bigcup_{\ell}{\cal
  K}_{\ell}$. Each of the wave vector shells ${\cal K}_{\ell}$ represents
an octave of wave numbers. The shell ${\cal K}_{0}$ describes 
the turbulent
motion of the large eddies which are of the order of the outer length scale
$L$. This shell is defined by $N$ wave vectors ${\bf k}^{(0)}_{i}$: ${\cal
  K}_{0}=\{{\bf k}^{(0)}_{i}:i=1,\dots,N\}$.  Starting with the generating
shell ${\cal K}_{0}$, the other shells ${\cal K}_{\ell}$ are found by a
successive rescaling of ${\cal K}_{0}$ with a scaling factor 2: ${\cal
  K}_{\ell}=2^{\ell} {\cal K}_{0}$. Thus each ${\cal K}_{\ell}$
consists of the $N$ scaled wave vectors $ 2^{\ell}{\bf k}_{i}^{(0)},\ 
i=1,\dots,N$. The shell ${\cal K}_{\ell}$ represents eddies at length scales
$r\sim 2^{-\ell}L$, i.e. to smaller and smaller eddies as the shell index
$\ell$ increases.
At scales $r \approx \eta$ the fluid motion is damped by
viscosity $\nu$, thus preventing the generation of infinitely 
small scales. Hence we only need to simulate shells ${\cal K}_{\ell},
\ell < \ell_{\nu}$, where $\ell_{\nu} \approx \log_2(L/\eta)$ 
is chosen such that the amplitudes in ${\cal K}_{\ell_{\nu}}$ are
effectively zero. In this representation the Navier--Stokes equation for
incompressible fluids reads for all ${\bf k}\in {\cal
  K}=\bigcup_{\ell=0}^{\ell_{\nu}}{\cal K}_{\ell}$:
\begin{mathletters}
  \label{incompNSeq}
  \begin{eqnarray}
    \frac{\partial}{\partial t}u_{i}({\bf k},t)&=&
       -\imath M_{ijk}({\bf k})\sum_{
      {\bf p},{\bf q}\in{\cal K}\atop {\bf k}={\bf p}+{\bf q}} u_{j}({\bf
      p},t)u_{k}({\bf q},t)-\nu k^{2}u_{i}({\bf k},t)+f_{i}({\bf
      k},t)\label{NSeq}\\
    {\bf k}\cdot {\bf u}({\bf k},t)&=&0\label{compress}\quad .
  \end{eqnarray}
\end{mathletters}
The coupling tensor $M_{ijk}({\bf k})=\left[k_{j}P_{ik}({\bf
  k})+k_{k}P_{ij}({\bf k})\right]/2$ with the projector $P_{ik}({\bf
  k})=\delta_{ik}-k_{i}k_{k}/k^{2}$ is symmetric in $j,k$ and $M_{ijk}({\bf
  k})=-M_{ijk}(-{\bf k})$. The inertial part of equation (\ref{NSeq}) consists
of all triadic interactions between modes with ${\bf k}={\bf p}+{\bf q}$.
They are the same as in the full Navier-Stokes equation for this triad.
The velocity field is driven by an external force ${\bf f}({\bf
  k},t)$ which simulates the energy input through a large-scale instability.

Within this approximation scheme the energy of a shell is 
\begin{equation}
  \label{energy}
  E_{\ell}(t)=\frac{1}{2}\sum_{{\bf k}\in{\cal K}_{\ell}} |{\bf u}({\bf
    k},t)|^{2}\quad ,
\end{equation}
and in the  absence of any viscous or external driving the 
total energy of the flow field
$E_{tot}(t)=\sum_{\ell=0}^{\ell_{\nu}} E_{\ell}(t)$ is conserved.
The choice of generating wave vectors ${\bf k}^{(0)}_i$ determines 
the possible triad interactions. This choice must at least guarantee 
energy transfer between shells and some mixing within a shell. In 
\cite{eggers91a,grossmann94a} different choices for 
wavenumber sets ${\cal K}_0$ are investigated. The larger the number 
$N$ of wave numbers, the more effective the energy transfer. Usually
one selects directions in ${\bf k}$-space to be distributed evenly 
over a sphere. However, there are different possibilities which 
change the relative importance of intra-shell versus inter-shell 
couplings. In this paper, we are going to investigate two different 
wave vector sets, with $N=26$ and $N=74$, which we call the small and the 
large wave vector set, respectively. In Fig. 1 a two-dimensional 
projection of both sets is plotted. The large wave vector set also
contains some next-to-nearest neighbor interactions between shells,
which we put to zero here, since they contribute little to the 
energy transfer. The small set allows 120 different interacting triads,
the large set 501 triads, 333 of which are between shells. 

\begin{figure}[H]
  \begin{center}
    \leavevmode
    \epsfsize=0.4 \textwidth
    \epsffile{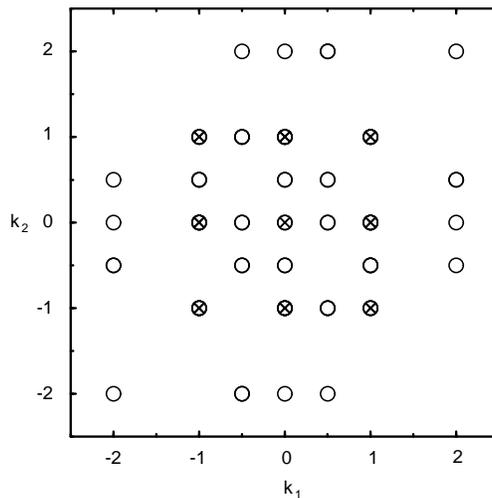}
    \caption{A two-dimensional projection of the $k$-vectors 
      in shell ${\cal K}_0$ for both the REWA models considered here.  
      The small set ($\times$) contains all vectors with -1, 0, and 1 as 
      components. The large set ($\bigcirc$), in addition, contains 
      combinations with $\pm1/2$ and $\pm 2$.
             }
\label{fig:sets}
  \end{center}
\end{figure}

\begin{figure}[H]
  \begin{center}
    \leavevmode
    \epsfsize=0.2 \textwidth
    \epsffile{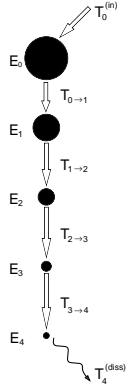}
    \caption{The structure of a local cascade. 
      Eddies of size $r\sim 2^{-\ell}L$ are
      represented by their total energy $E_{\ell}$. 
      Only modes of
      neighboring shells interact, leading to a local energy
      transfer $T_{\ell\rightarrow\ell+1}(t)$. The cascade is driven by
      injecting energy into the largest scale with rate
      $T_{0}^{(in)}(t)$. The turbulent motion is damped by viscous dissipation
      at a rate 
      $T_{\ell}^{(diss)}(t)$. 
             }
\label{fig:modelstructure}
  \end{center}
\end{figure}

Since in the models we consider energy transfer is purely local,
the shell energies $E_{\ell}(t),\ \ell=0,\dots,\ell_{\nu}$ only change 
in response to energy influx $T_{\ell-1\rightarrow \ell}$ from above
and energy outflux $T_{\ell\rightarrow \ell+1}$ to the lower 
shell. In addition, there is a rate of viscous dissipation
$T_{\ell}^{(diss)}(t)$ which is concentrated on small scales, 
and a rate of energy input $T_{0}^{in}(t)$, which feeds
the top level only, cf. Fig.\ref{fig:modelstructure}.

From equation (\ref{NSeq}) we find an energy balance equation which governs
the time evolution of the shell energies $E_{\ell}(t)$
\begin{equation}
  \label{energyconservationlaw}
  \frac{d}{dt} E_{\ell}(t)=T_{\ell-1\rightarrow \ell}(t)-T_{\ell\rightarrow
    \ell+1}(t)+T_{\ell}^{(diss)}(t)+T_{0}^{(in)}(t)\delta_{\ell0}\quad .
\end{equation}
The different transfer terms are found to be
\begin{mathletters}
\label{FWtransfer}
\begin{eqnarray}
  T_{\ell\rightarrow \ell+1}(t) &=& 
       2\imath \sum_{\triangle^{(\ell+1)}_{(\ell)}}
  M_{ijk}({\bf k}) u_{i}^{*}({\bf k},t)u_{j}({\bf p},t)u_{k}({\bf q},t)
  \label{FWta}\\   
  T_{0}^{(in)}(t) &=& \sum_{{\bf k}\in{\cal K}_{0}} \mbox{\rm Re}\left({\bf
    u}^{*}({\bf k},t)\cdot{\bf f}({\bf k},t)\right) \label{FWtb}\\ 
  T_{\ell}^{(diss)}(t) &=& -\nu \sum_{{\bf k}\in{\cal K}_{\ell}} k^{2}|{\bf
    u}({\bf k},t)|^{2} \label{FWtc}\quad .
\end{eqnarray}
\end{mathletters}
In equation (\ref{FWta}) $\sum_{\triangle^{(\ell+1)}_{(\ell)}}$ 
indicates the summation over all
next--neighbor triads ${\bf k}={\bf p}+{\bf q}$ with ${\bf k}\in {\cal
  K}_{\ell},\ {\bf p}\in{\cal K}_{\ell+1}$ and ${\bf q}\in{\cal
  K}_{\ell}\bigcup{\cal K}_{\ell+1}$.

The driving force ${\bf f}({\bf k},t)$ is assumed to act only on the 
largest scales, and controls the rate of energy input
$T_{0}^{(in)}(t)$. As in Reference \onlinecite{eggers91a}
we choose ${\bf f}({\bf k},t)$ to ensure constant energy input 
$T^{(in)}_0 = \epsilon$ : 

\begin{mathletters}
    \label{force}
    \begin{eqnarray}
      {\bf f}({\bf k},t)&=&\frac{\epsilon {\bf u}({\bf k},t)}{2N|{\bf
          u}({\bf k},t)|^{2}} \enspace \mbox{for all } {\bf k}\in{\cal
        K}_0\\ {\bf f}({\bf k},t)&=&0 \enspace \mbox{for all } {\bf
        k}\not\in{\cal K}_0 \quad .
    \end{eqnarray}
\end{mathletters}

\begin{figure}[H]
  \begin{center}
    \leavevmode
    \epsfsize=0.4 \textwidth
    \epsffile{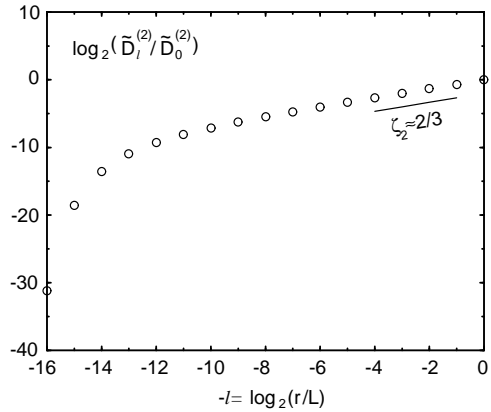}
    \caption{Log--log--plot of the structure function 
      $\tilde{D}_{\ell}^{(2)}=\left<E_{\ell}\right>$ versus level number
      for the small cascade. At large scales, where the influence of 
      dissipation is negligible, classical scaling is observed. 
      At small scales the
      turbulent motion is damped by viscosity. 
      The Reynolds number is $Re = 4.2\cdot10^5$.
             }
    \label{fig:scaling}
  \end{center}
\end{figure}

\begin{figure}[H]
  \begin{center}
    \leavevmode
    \epsfsize=0.4 \textwidth
    \epsffile{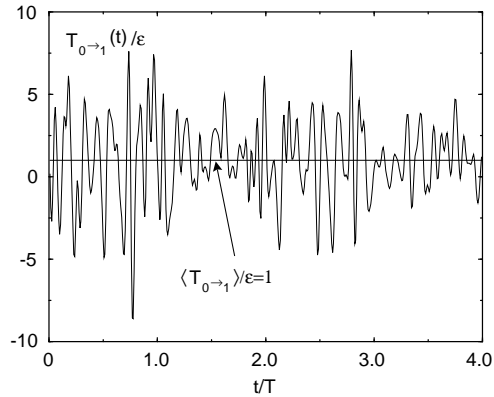}
    \caption{Time evolution of the energy transfer $T_{0\rightarrow
        1}(t)$ from shell ${\cal K}_{0}$ to shell ${\cal K}_{1}$. 
       Typical excursions are large compared with the mean value 
       $\left<T_{0\rightarrow
        1}\right>=\epsilon$. 
       The time is given in units of
      $T = \left<E_{0}\right>/\epsilon$, where $\left<E_{0}\right>$ 
      is the mean energy of the top level.}
    \label{fig:transfer}
  \end{center}
\end{figure}

This leads to a
stationary cascade whose statistical properties are governed by the 
complicated chaotic dynamics of the nonlinear mode interactions. 
Owing to energy conservation, viscous dissipation equals the energy 
input on average. The Reynolds number is given by 
\begin{equation}
  \label{re}
    Re = \frac{L U}{\nu} = \frac{\epsilon L^2}{\left< E_0 \right>\nu} \quad,
\end{equation}
since $T = \left<E_0\right> / \epsilon$ sets a typical turnover 
time scale of the energy on the highest level. We believe 
$T$ to be of particular relevance, since the large-scale fluctuations
of the energy will turn out to be responsible for the intermittent 
behavior we are interested in. In \cite{grossmann94a}, for example, 
time is measured in units of $L^{2/3} \epsilon^{-1/3}$, which 
typically comes out to be 1/10th of the turnover time of the energy $T$.
As we are going to see below, this is rather a measure 
of the turnover times of the individual Fourier modes. 
Figure \ref{fig:scaling} shows the scaling of the mean energy 
in a log-log plot at a Reynolds number of $4.2 \cdot 10^5$.
The inertial range extends over three decades, where a power
law very close to the prediction of classical scaling is seen. 
Below the 10th level the energies drop sharply due to viscous 
dissipation. In Section \ref{sec:equilib} we are going to turn 
our attention to the small corrections to 2/3-scaling, hardly 
visible in Fig. \ref{fig:scaling}. Still, there are considerable 
fluctuations in this model, as evidenced by the plot of the energy 
transfer in Fig. \ref{fig:transfer}. Typical excursions from the 
average, which is normalized to one, are quite large. Ultimately, 
these fluctuations are responsible for the intermittency 
corrections we are going to observe. 

\begin{figure}[H]
  \begin{center}
    \leavevmode
    \epsfsize=0.4 \textwidth
    \epsffile{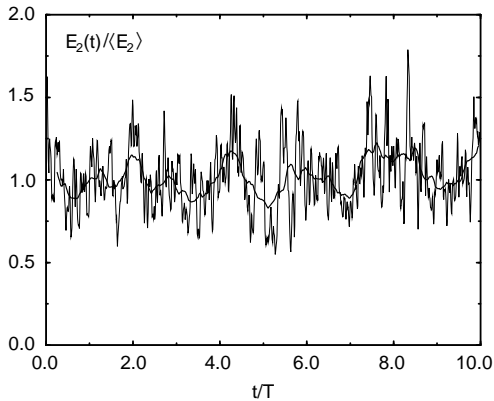}
    \caption{Energy of a shell ($\ell=2$) as a function of time. 
       The rapid fluctuations come 
      from the motion of individual Fourier modes. 
      A much longer time scale is revealed by performing a 
      floating average over one turnover time of
      the second level (bold line).
      The time is given in units of $T = \left<E_{0}\right>/\epsilon$.}
    \label{fig:energy}
  \end{center}
\end{figure}

Thus within the REWA--cascade we
are able to numerically analyze the influence of fluctuations
on the stationary statistical properties of a cascade with local energy
transfer on the basis of the Navier--Stokes
equation. In Fig. \ref{fig:energy} we plot the time evolution 
of the energy on the second level of the cascade. One observes

short-scale fluctuations, which result from the motion of individual 
Fourier modes within one cascade level. However, performing a floating 
average reveals a {\it second} time scale, which is of the same order as the 
turnover time of the top level. As we are going to see in Section
\ref{sec:fdt}, this disparity of time scales is even  
more pronounced on lower levels. The physics idea is the same as 
in the microscopic foundation of hydrodynamics,
where conserved quantities are assumed to move on much slower time scales
than individual particles.
This motivates us to consider the energy as the only dynamical variable
of each shell, and to represent the rapid fluctuations of 
Fig. \ref{fig:energy} by a white-noise Langevin force. In this approximation
we still hope to capture the rare, large-scale events characteristic 
of intermittent fluctuations, since the conserved 
quantity is the ``slow'' variable of the system. Similar ideas have also 
been advanced for the conservative dynamics of a non-equilibrium
statistical mechanical system \cite{spohn}.
  
\subsection{The Langevin--cascade}

In this model we take a phenomenological view of the process of energy 
transfer. The chaotic dynamics of the REWA cascade is modeled by a stochastic 
equation. We make sure to include the main physical features of 
energy conservation and local coupling. In particular,
the dynamics is simple enough to allow for analytical insight into
the effects of fluctuating energy transfer \cite{eggers94}.

As in the preceding REWA--cascade, the turbulent flow field is 
described by a sequence of eddies decaying successively
(Figure \ref{fig:modelstructure}). The eddies at
length scales $r\sim 2^{-\ell}L$ are represented by their energy
$E_{\ell}(t)$.
As before we restrict ourselves to local energy transfer,
and thus the time evolution of the
shell energies $E_{\ell}(t)$ is governed by equation
(\ref{energyconservationlaw}). The crucial step is of course to choose 
an appropriate energy transfer $T_{\ell\rightarrow\ell+1}(t)$. 
For simplicity, we restrict ourselves to a Langevin process with a white 
noise force. Thus the local transfer $T_{\ell\rightarrow \ell+1}(t)$ is
split into a deterministic and a stochastic part $T_{\ell\rightarrow
  \ell+1}(t) = T_{\ell\rightarrow \ell+1}^{(det)}(t)+
T_{\ell\rightarrow\ell+1}^{(stoch)}(t)$ where both
parts should depend only on the local length scale $2^{-\ell}L$ and
the neighboring energies $E_{\ell}$ and $E_{\ell+1}$. The most 
general form dimensionally consistent with this has been given in 
\cite{eggers94}. For simplicity, here we restrict ourselves to the specific
form
\begin{mathletters}
  \label{Letrans}
  \begin{eqnarray}
    T_{\ell\rightarrow \ell+1}^{(det)}(t)&=& D
    \frac{2^{\ell}}{L}\left(E_{\ell}^{3/2}(t)-E_{\ell+1}^{3/2}(t)\right)
    \label{Letransa}\\ 
    T_{\ell\rightarrow \ell+1}^{(stoch)}(t)&=& R
    \left(\frac{2^{(\ell+1)}}{L}\right)^{1/2}
      E_{\ell}^{5/8}(t)E_{\ell+1}(t)^{5/8}
      \xi_{\ell+1}(t)\label{Letransb}\\
      T_{\ell}^{(in)}(t)&=& \epsilon
    \delta_{\ell 0} \label{Letransc}\\
     T^{(diss}_{\ell}(t) &=& 
        -\nu(2^{-\ell}L)^{-2} E_{\ell} \quad. \label{Letransd}
  \end{eqnarray}
\end{mathletters}
The white noise is represented by 
$\xi_{\ell}$, i.e. $\left<\xi_{\ell}(t)\right>=0$ and
$\left<\xi_{\ell}(t)\xi_{\ell'}(t')\right>=2\delta_{\ell\ell'}\delta(t-t')$.
We use Ito's \cite{gardiner83} definition in equation (\ref{Letransb}). 
To understand the dimensions appearing in (\ref{Letrans}), note 
that $u_{\ell} \sim E_{\ell}^{1/2}$ is a local velocity scale and 
$k \sim 2^{\ell}/L$ is a wavenumber. Thus (\ref{Letransa}) 
dimensionally represents the energy transfer (\ref{FWta}).
In (\ref{Letransb}) the powers are different, since $\xi$ carries
an additional dimension of $1/\mbox{time}^{1/2}$. 
It follows from (\ref{Letransa}) that the sign of the deterministic 
energy transfer depends on which of the neighboring energies 
$E_{\ell}$ or $E_{\ell+1}$ are greater. If for example $E_{\ell}$ 
is larger, $T^{(det)}_{\ell\rightarrow\ell+1}(t)$ is positive, 
depleting $E_{\ell}$ in favor of $E_{\ell+1}$. Hence the 
deterministic part tends to equilibrate the energy among the 
shells. The stochastic part, on the other hand, is symmetric with 
respect to the two levels $\ell$ and $\ell+1$. This reflects 
our expectation that in equilibrium it is equally probable for 
energy to be scattered up or down the cascade. 

The combined effect of (\ref{Letransa}) and (\ref{Letransb}) is 
that without driving, energies fluctuate around a common mean value. 
This equipartition of energy in equilibrium is precisely 
what has been predicted on the basis of the Navier-Stokes 
equation \cite{kraichnan73,orszag93}. 
The only free parameters appearing in the transfer are thus the 
amplitudes $D$ and $R$. If $R$ is put to zero, the motion is 
purely deterministic, and one obtains the simple solution 
\begin{equation}
  \label{statenerg}
  E_{\ell}^{(0)}=C 2^{-(2/3)\ell}\ \mbox{with}\ 
           C=\left(\frac{2\epsilon L}{D}\right)^{2/3}\quad .
\end{equation}
This corresponds to a classical Kolmogorov solution with no 
fluctuations in the transfer. The amplitude $D$ of the deterministic
part is a measure of the effectiveness of energy transfer. On the other 
hand $R$ measures the size of fluctuations. In \cite{eggers94}
it is shown that a finite $R$ necessarily leads to intermittency 
corrections in the exponents. In the next section we are going to 
determine the model parameters for the two REWA cascades we are 
considering.


\section{Determination of model parameters}
\label{sec:equilib}

The aim of this section is to explain the significance of the 
parameters appearing in the Langevin model. We show that only the 
combination $R/D^{1/2}$ determines the nonequilibrium fluctuations
in the inertial range. But it is the same combination which also 
sets the width of the {\it equilibrium} distribution of energies, 
if the chain of shells is not driven. Thus we are able to fix all
the parameters necessary to describe the nonequilibrium state 
solely by measuring equilibrium properties. 

The physical parameters of the turbulent cascade are the length scale 
$L$ of the highest level, the rate of energy input $\epsilon$, and
the viscosity $\nu$. The properties of the energy transfer are
determined by the dimensionless strength of the deterministic 
part $D$ and of the stochastic part $R$. In the previous section we have 
seen that $E^{(0)}_0 = (2\epsilon L/D)^{2/3}$ sets an energy scale 
for the highest level, and thus $T = E^{(0)}_0 / \epsilon$ is a 
time scale. Both scales can be used to non-dimensionalize 
(\ref{Letrans}), giving
\begin{mathletters}
\label{nondimLetrans}
\begin{eqnarray}
  \hat{T}_{\ell\rightarrow \ell+1}^{(det)}(t) &=& 
  2^{\ell+1}\left(\hat{E}_{\ell}^{3/2}(t)-\hat{E}_{\ell+1}^{3/2}(t)\right)
    \label{nondimLetransa}\\ 
    \hat{T}_{\ell\rightarrow \ell+1}^{(stoch)}(t)&=& \frac{R}{D^{1/2}}
    2^{(\ell+2)/2}\hat{E}_{\ell}^{5/8}(t)\hat{E}_{\ell+1}(t)^{5/8}
      \hat{\xi}_{\ell+1}(t)\label{nondimLetransb}\\
      \hat{T}_{0}^{(in)}(t)&=& 1
     \label{nondimLetransc} \\
      \hat{T}_{\ell}^{(diss)}(t)&=& -2^{2\ell}
      (Re)^{-1} \hat{E}_{\ell} \quad .\label{nondimLetransd} \\
\end{eqnarray}
\end{mathletters}
Hence $R/D^{1/2}$ is the only parameter characterizing the inertial
range dynamics, which can be understood as follows: At any level $\ell$,
a time scale of deterministic transport is set by $\tau_D \sim 
E_{\ell}/\epsilon \sim E_{\ell}^{-1/2} r /D$, where $E_{\ell}$ 
is a typical energy scale and $r = 2^{-\ell} L$. This deterministic 
transport competes with diffusion of energy introduced by the 
stochastic part of the energy transfer. Namely a time scale over 
which the energy $E_{\ell}$ can diffuse away is given by 
$\tau_R \sim (E_{\ell}/A)^2$, where $A = R \epsilon r^{-1/2} E_{\ell}^{3/4}$ 
is the amplitude of the noise term. 
Plugging in the expression from (\ref{Letransb}) one ends up with 
$\tau_R \sim r E_{\ell}^{-1/2} / R^2$, and hence 
\[
\frac{\tau_R}{\tau_D} \sim \frac{D}{R^2} .
\]
In particular, the relative importance of deterministic transport 
and diffusion is independent of the level, and only depends on 
the combination of $D$ and $R$ given above. 

\begin{figure}[H]
  \begin{center}
    \leavevmode
    \epsfsize=0.4 \textwidth
    \epsffile{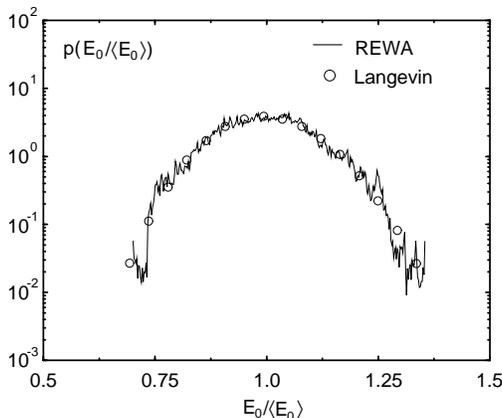}
    \caption{Equilibrium distribution of the energy in the top level
      of the large REWA--cascade ($N=74$), consisting of 7 shells.
      By adjusting the only parameter $R/D^{1/2}$, the distribution of the 
      Langevin cascade can be made to agree within statistical error. 
      The results for $R/D^{1/2}$ for both wave vector sets are
      given in (\protect\ref{rdparam}).
              }
    \label{fig:equilib}
  \end{center}
\end{figure}

We are now left to determine $R/D^{1/2}$ from an experiment which 
is independent of the nonequilibrium state. To that end we 
consider a long chain of shell elements, for which energy input
as well as viscous dissipation has been turned off ($\epsilon = 
\nu = 0$). As a result, the energy will perform fluctuations around 
some mean value $\left<E_{\ell}\right>$. Using a similar argument as 
above, the probability distribution $p(E_{\ell} / \left<E_{\ell}\right>)$
for the shell energies will only depend on $R/D^{1/2}$ for the 
Langevin model. Figure \ref{fig:equilib} shows the probability distribution
for one level of a cascade with 7 shells. Except for some end effects
at the lowest level, the distribution turns out to be level-independent.
The large REWA cascade is compared 

with a simulation of the Langevin cascade. Once $R/D^{1/2}$ is adjusted, 
there is an almost perfect match between the two models. This shows
that the simple stochastic dynamics proposed here models the equilibrium 
distribution of the chaotic fluctuations of Fourier modes very well. 
Most importantly, we have determined the only adjustable parameter 
of the energy transfer of the Langevin cascade corresponding to the two 
model systems we are considering: 
\begin{mathletters}
\label{rdparam}
\begin{eqnarray}
\frac{R}{D^{1/2}} &= 1.729\cdot 10^{-1} & \mbox{ for the small cascade}\;
(N=26),\label{rdparama}\\
\frac{R}{D^{1/2}} &= 9.968\cdot 10^{-2} & \mbox{ for the large cascade}\;
                        (N=74) \label{rdparamb}\quad .
\end{eqnarray}
\end{mathletters}
As seen from (\ref{rdparam}), the REWA cascade with a larger number of 
modes has smaller fluctuations. This is not surprising, since large 
excursions of the energy correspond to a coherent motion of the 
individual Fourier modes. The larger the number of modes, the harder this is 
to achieve, since the random motion of individual modes tends to 
destroy correlations. 

In the next section, the fit (\ref{rdparama}),(\ref{rdparamb}) will be
used to compare nonequilibrium properties in the inertial range. 
All information about possible differences between the small and 
the large cascade has been condensed into a single number.



\section{Intermittency Corrections}
\label{sec:fdt}

We now turn to the inertial range fluctuations of the two 
cascade models. To allow for a direct comparison, we focus
on the scaling of the energies $E_{\ell}$ of one shell. 
The analogue of the moments of the velocity field usually 
considered in turbulence are the structure functions based
on the energy 
\begin{equation}
  \label{scaling}
  \tilde{D}^{(p)}_{\ell} =
       \left<E_{\ell}^{p/2}\right>\sim 2^{-\zeta_{p}\ell}\quad .
\end{equation}
We concentrate on the small corrections $\delta\zeta_p$ to the 
scaling exponents as predicted on dimensional grounds. As seen 
in Fig. \ref{fig:scaling}, these corrections are extremely 
small, so that no significant deviation from classical scaling
is seen on the scale of the figure. 

In order to know what to expect, we use the result of 
Ref. \cite{eggers94} where we have computed the scaling behavior 
of a general class of stochastic models in a perturbation expansion.
To lowest order, the exponent corrections are given by the 
quadratic dependence 
\begin{equation}
  \label{intermittcorr}
  \delta\zeta_{p}=-\mu \frac{p}{18}(p-3)\quad .
\end{equation}
Plugging the specific form of the energy transfer (\ref{nondimLetransa}),
(\ref{nondimLetransb}) into the formulae given in \cite{eggers94}, we find 
\begin{equation}\label{prediction}
  \mu = 0.42 \left(\frac{R}{D^{1/2}}\right)^{2}\quad.
\end{equation}
For the small cascade this leads to $\mu = 0.013$, which is about 
1/10 of the experimental value accepted for three-dimensional 
turbulence \cite{anselmet84}. Therefore, it is more than ordinarily 
difficult to measure the scaling exponents with sufficient 
accuracy to obtain significant answers for the deviations from
classical scaling. On the other hand, the number of modes in a 
REWA cascade being greatly reduced as compared with the full 
Navier-Stokes equation, we are able to simulate cascades with up to 
17 levels, corresponding to 5 decades in scale. 

\begin{figure}[H]
  \begin{center}
    \leavevmode
    \epsfsize=0.4 \textwidth
    \epsffile{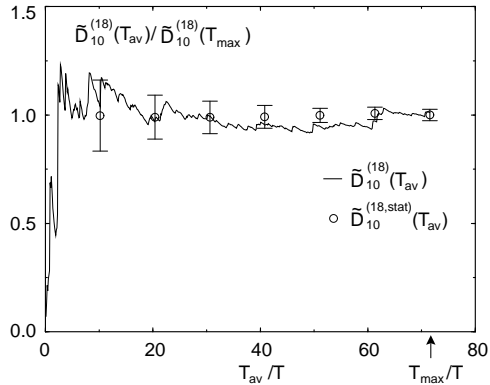}
    \caption{Convergence of the structure function
      $\tilde{D}_{\ell}^{(18)}=\left<E_{\ell}^{9}\right>$, on the 
      11th step of the small REWA cascade. We show the evolution of
      $\tilde{D}_{\ell}^{(18)}$ as function of the averaging time $T_{av}$
      in comparison with the stationary values 
      $\tilde{D}_{\ell}^{(18,stat)}$, which are calculated by integrating 
      over all initial values in the data set. The error bars give the
      statistical error of the data as estimated by 
      (\protect{\ref{variance}}). 
      The averaging
      time $T_{av}$ is given in units of
      $T = \left<E_{0}\right>/\epsilon$.}
    \label{fig:convergence}
  \end{center}
\end{figure}

To obtain reliable results within the accuracy needed, it is essential 
to disentangle statistical errors from systematic errors, introduced
through finite-size effects or viscous damping. To that end we individually
assign statistical errors to every average taken. Figure \ref{fig:convergence}

shows the convergence of the 9th moment of the energy on the 11th step 
of the small REWA cascade, which is the lowest level relevant to our 
fits. We plot the mean value, averaged over the time given. Evidently 
the fluctuations of this mean value get smaller as the averaging
time $T_{av}$ is increased. To obtain a 
quantitative measure of the uncertainty of the $T$-average of some quantity 
$x$, we also consider the ensemble of averages 
obtained with different initial conditions. The result for the 
mean value and variance of this ensemble average is plotted 
as circles with error bars in Fig. \ref{fig:convergence}, where 
$E_{\ell}^9$ takes the place of $x$. The variance is 
found equivalently as an integral over the temporal correlation 
function \cite{tennekes80}: 
\begin{equation}
  \label{variance}
   \sigma_x^2(T_{av}) = \frac{1}{T_{av}^2} \int^{T_{av}}_0
    \int^{T_{av}}_0 \left[\left<x(\tau_1)x(\tau_2)\right> - 
        \left<x\right>^2\right] d\tau_1d\tau_2 \quad.
\end{equation}
This variance $\sigma(T_{av})$ is seen to give a reasonable approximation 
to the fluctuations of the temporal average. The variance
for the largest averaging time available has been taken as the statistical 
error of a measured temporal average. 

\begin{figure}[H]
  \begin{center}
    \leavevmode
    \epsfsize=0.4 \textwidth
    \epsffile{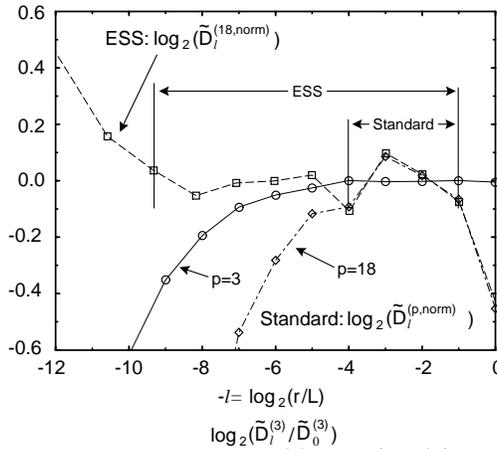}
    \caption{Scaling of the structure functions
      $\tilde{D}_{\ell}^{(p)}=\left<E_{\ell}^{p/2}\right>$ with
      $p=3$ ($\bigcirc$) and $p=18$ ($\Diamond$) versus level number
      (Standard) in comparison with
      extended self similarity (ESS) scaling of 
       $\tilde{D}_{\ell}^{(18)}$ ($\Box$) versus 
       $\tilde{D}_{\ell}^{(3)}/\tilde{D}_{0}^{(3)}$. The structure functions
       are normalized by their power law fits:
        $\tilde{D}_{\ell}^{(p,norm)} = 
        \tilde{D}_{\ell}^{(p)}/(b^{(p)}2^{-\zeta_{p}\ell})$
        in the standard plot and
        $\tilde{D}_{\ell}^{(p,norm)} = 
        \tilde{D}_{\ell}^{(p)}/(A^{(p)}(\tilde{D}_{\ell}^{(3)}/
        \tilde{D}_{0}^{(3)})^{\zeta_{p}/\zeta_{3}})$
        in the ESS plot. Refer to Table 
        \protect\ref{tab:fw-exponents} for the values
        of the exponents. In the ESS plot the scaling range is 
        more than doubled.
        The numerical calculation is
        performed for the small REWA cascade.
                  }
    \label{fig:ess}
  \end{center}
\end{figure}

Next we consider systematic errors in the computation of the scaling 
exponents. Deviations from power law scaling are expected to occur
on both ends of the cascade and have been studied extensively 
\cite{eggers91a,grossmann92,grossmann94a}. First,
fluctuations on the highest level are suppressed, since there is no 
coupling to a higher level, but deterministic energy input instead. 
Second, in the viscous subrange the energy is increasingly depleted 
by viscosity, leading to even more drastic effects on the spectra. 
Both effects are strongest for the highest moments, which are most 
sensitive to large fluctuations. The 18th order structure function for a 
typical run of the small cascade is plotted as diamonds in 
Fig. \ref{fig:ess}. The average on the highest level is considerably 
lower than expected from scaling, as fluctuations are suppressed. 
The cascade only gradually recovers from this suppression, which leads 
to a faster rise in the level of fluctuations and thus to a decrease in the 
local scaling exponent as reported in \cite{grossmann94a}. 
Note that we show a ``scatter-plot'' of the data around the power 
law $\tilde{D}^{(p)}_{\ell} \sim 2^{-\ell\zeta_p}$, which represents 
our best fit. Thus deviations from this power law are hugely exaggerated 
and would not be visible on a customary log-log-plot.
The absolute variation of $\tilde{D}^{(18)}_{\ell}$ over the range of the 
plot is 21 orders of magnitude. 
Owing to the extreme sensitivity of our plot, viscous damping is visible 
below the fourth level, even for a cascade with 17 shells, as seen in 
Fig. \ref{fig:ess}. This restricts the inertial range to four levels. 
We employ two different strategies to improve on this situation:
(i) We plot the higher order structure functions against the third order 
structure function, as suggested by the ESS method. (ii) We eliminate
viscosity altogether by introducing a fluctuating eddy viscosity instead.

It is known \cite{benzi93a,benzi93b} that extended self-similarity
leads to a very considerable 
improvement of the scaling of experimental data. Without any viscous 
corrections, the scaling behavior is unaffected, as seen in 
Fig. \ref{fig:ess} in the fitting range marked ``standard''.
Below the forth level, however, the third order structure function
suffers viscous damping quite similar to that affecting the other 
structure functions.  
\begin{table}[bt]
  \begin{center}
    \leavevmode
    \begin{tabular}[t]{c*{2}{d@{${}\cdot{}$}l@{${}\pm{}$}d@{${}\cdot{}$}l}
        *{1}{d@{${}\cdot{}$}l}} 
      p& \multicolumn{4}{c}{$\delta\zeta_p$ (ESS)}
       & \multicolumn{4}{c}{$\delta\zeta_p$ (EV)} 
       & \multicolumn{2}{c}{$\delta\zeta_p$ (Theory)} 
      \\ \tableline
      2&1.5&$10^{-3}$&1.5&$10^{-3}$
       &1.4&$10^{-3}$&2&$10^{-4}$
       &1.4&$10^{-3}$
      \\ \tableline
      6&$-$1.3&$10^{-2}$&5&$10^{-3}$
       &$-$1.3&$10^{-2}$&1&$10^{-3}$
       &$-$1.3&$10^{-2}$
      \\ \tableline
     12&$-$7.5&$10^{-2}$&1&$10^{-2}$
       &$-$7.4&$10^{-2}$&2&$10^{-3}$
       &$-$7.6&$10^{-2}$
      \\ \tableline
     18&$-$1.8&$10^{-1}$&2&$10^{-2}$
       &$-$1.9&$10^{-1}$&1&$10^{-2}$
       &$-$1.9&$10^{-1}$
    \end{tabular}
  \end{center}
  \caption{Correction exponents of the
 structure functions $\tilde{D}_{\ell}^{(p)}$ for $p=2,6,12$ and $18$ for the 
small REWA cascade. The exponents in column ESS are determined by 
plotting $\tilde{D}^{(p)}_{\ell}$ versus $\tilde{D}^{(3)}_{\ell}$ 
as suggested by extended self-similarity. To obtain the values 
in column EV, viscous effects have been removed entirely by 
introducing a fluctuating eddy viscosity. In the last column the 
prediction based on the Langevin model is given.
         }
  \label{tab:fw-exponents}
\end{table}
Thus by plotting $\tilde{D}^{(3)}_{\ell}$ on the
abscissa, both effects are hoped to largely cancel each other. This 
is indeed true for the REWA cascade as well, where
the scaling range has been extended to almost three decades. In 
Table \ref{tab:fw-exponents} we have compiled various scaling 
exponents for the small cascade. Except for the smallest moment, 
highly significant corrections to the scaling exponents are measured.
The errors are based on a least squares fit with weighted 
averages \cite{press92}, based on the statistical errors as 
explained above. 
To make sure the results do not depend on our choice of the inertial
range, we also made fits in ranges other than the one marked ``ESS'' 
in Fig. \ref{fig:ess}. Namely, we variously shifted the fitting range
by one level up or down, or took an additional level into account
on either end. In each case, the values of the exponents were within
the errors given in Table \ref{tab:fw-exponents}. 
Typical averaging times were 100 turnover times 
of the largest scale. This is about 10 times as long as in 
\cite{grossmann94a}, as we base the turnover time on the time scale
of the energy.
For comparison, we also supply the value of the exponent correction
as calculated from (\ref{intermittcorr}) and (\ref{prediction})
on the basis of the known value of $R/D^{1/2}$, given in 
(\ref{rdparam}). There is no adjustable parameter in this comparison
with the theoretical prediction, since the noise strength was 
determined solely from the {\it equilibrium} fluctuations of
the REWA cascade. The excellent agreement found for all exponents 
makes us confident that finite size corrections and viscous effects
have been successfully eliminated, and we are measuring genuine 
properties of the inertial range. 

\begin{figure}[H]
  \begin{center}
    \leavevmode
    \epsfsize=0.4 \textwidth
    \epsffile{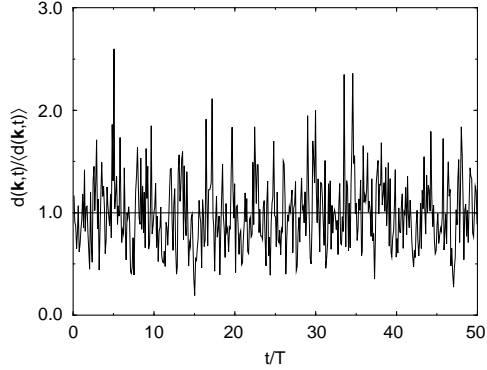}
    \caption{Time evolution of the
      amplitude $d({\bf k},t)$ of the eddy damping 
      (\protect\ref{eddyviscosity}) for ${\bf
        k}=2^{l_{max}}(1,1,1)$. The fluctuations 
       mirror the highly
      intermittent behavior of the velocity modes on the 14th level. The
      numerical calculation is performed with the small wave
      vector set. Time is given in units of
      $T = \left<E_{0}\right>/\epsilon$.} 
    \label{fig:eddyvisc}
  \end{center}
\end{figure}

However, since there is no theory demonstrating that the ESS method
leads to the correct inertial range scaling behavior, we have nevertheless 
attempted to eliminate viscous effects by a second, and completely 
independent method. This was done by putting $\nu = 0$, and instead 
draining energy from the lowest shell ${\cal K}_{\ell_{max}}$ using
an eddy viscosity \cite{kerr78,eggers91a}. In order to mimic 
inertial energy transfer into the subgrid shells ${\cal K}_{\ell},
\ell > \ell_{max}$, we preserve the coupling structure of the equations as 
much as possible. We add a term 
\begin{equation}
  \label{eddyviscosity}
  {\bf D}({\bf k},t)=\left\{
    \begin{array}{r@{\quad:\quad}l}
      -d({\bf k},t)\, 2^{\ell_{max}}|{\bf u}({\bf k},t)|{\bf u}({\bf k},t) &
      {\bf k}\in{\cal K}_{(out)}\\ 0 & {\bf k}\not\in{\cal K}_{(out)}
    \end{array}\right. \quad 
\end{equation}
to the inviscid Navier-Stokes equation (\ref{NSeq}). The cut-off
shell ${\cal K}_{(out)} \subset {\cal K}_{\ell_{max}}$ contains all 
the wave vectors of ${\cal K}_{\ell_{max}}$ which interact directly 
with modes of the shell which is not resolved. The problem of this
procedure is that (i) the fluctuations in the subgrid scales are 
not accounted for, and (ii) the amplitude to choose for $d({\bf k},t)$
is not known. Both problems are addressed by using a method inspired 
by the work in \cite{domaradzki95}. The idea is to adjust 
$d({\bf k},t)$ such that 
\begin{equation}
  \label{ev}
  \left<|{\bf u}({\bf k})|^{3}\right> = 
     \frac{1}{2}\left<|{\bf u}({\bf k'})|^{3}\right>\mbox{
    for } {\bf k}\in{\cal K}_{\ell_{max}},\, 
          {\bf k'}=\frac{1}{2}{\bf k}\quad,
\end{equation}
which is valid exactly in a perfectly scale-invariant cascade 
\cite{eggers91a}, analogous to the Kolmogorov structure 
equation \cite{kolmogorov41}. The wave vector ${\bf k}' \in
{\cal K}_{\ell_{max}-1}$ in (\ref{ev}) is a scaled copy of 
${\bf k} \in {\cal K}_{\ell_{max}}$. In the long time limit,
each $d({\bf k},t)$ will converge to some average value, which
is determined self-consistently from the dynamics of the cascade. 
But in that case the {\it fluctuations} of the subgrid scales would
be missing. Therefore, we instead took the averages in (\ref{ev})
over just 10 turnover times of the last resolved shell 
${\cal K}_{\ell_{max}}$. When $\left<|{\bf u}({\bf k})|^3\right> / 
\left<|{\bf u}({\bf k'})|^3\right>$ was larger than 1/2, we 
increased $d({\bf k},t)$ by 5 \%, otherwise $d({\bf k},t)$ 
was decreased by the same percentage. Thus all the $d({\bf k},t)$ 
reflect the fluctuations occurring at the end of the cascade, as seen
in Fig. \ref{fig:eddyvisc} for one of the amplitudes of the small 
cascade. The average of $d({\bf k},t)$ becomes stationary in the 
long time limit, but fluctuations are considerable, as expected 
in an intermittent cascade. 

\begin{figure}[H]
  \begin{center}
    \leavevmode
    \epsfsize=0.4 \textwidth
    \epsffile{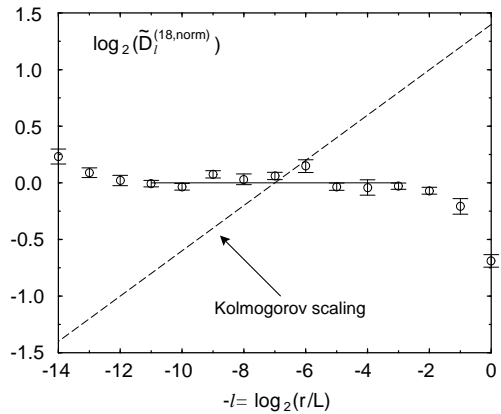}
    \caption{Anomalous scaling of the structure function
      $\tilde{D}_{\ell}^{(18)},\ p=18$ for the small REWA cascade,
      with the eddy viscosity (\protect{\ref{eddyviscosity}}) 
      on the lowest level. We
      show the scatter of the data around the fitted power law by plotting
      $\tilde{D}_{\ell}^{(18,norm)} = 
       \tilde{D}_{\ell}^{(18)}/(b^{(18)}2^{-\zeta_{18}\ell})$. 
       Classical Kolmogorov scaling is shown for comparison.
                 }
    \label{fig:fitquality}
  \end{center}
\end{figure}

Figure \ref{fig:fitquality} shows the 9th moments of the energy
as in Fig. \ref{fig:ess}, but for the eddy-damped cascade. End
effects at the small scales are very small, less substantial 
than on the largest scales. Also shown are the error bars resulting
from the statistical estimate described before. 
The result of the 
fit over the scaling range indicated is given in Table 
\ref{tab:fw-exponents} for various moments. Again, the range of our fit
was also varied, and the results are consistent with the errors given. 
The values of the 
exponent corrections are in excellent agreement with the values obtained 
from ESS. Furthermore, the error is even less than before, giving very 
significant deviations from classical scaling. This is also seen
from the dashed line in Fig. \ref{fig:fitquality}, which represents
classical Kolmogorov scaling. Finally it should be noted that
even for the long averaging times we use, some imbalances in the cascade
remain, which make $\tilde{D}^{(3)}_{\ell}$ deviate from its 
exact power law behavior of $2^{-\ell}$. These deviations decrease
in time and turn out not to be significant for many of our runs. 
However, we decided to normalize our results so as to substract 
remaining imbalances. So strictly speaking the values given for the 
correction exponents in Table \ref{tab:fw-exponents} are 
\[
  \delta\zeta_p = \frac{\zeta_p}{\zeta_3} - \frac{p}{3}  \quad.
\]
The same procedure has been adopted in previous work on the subject
\cite{grossmann94a}. 

Since the best error estimates are obtained by using an eddy-damped cascade, 
we are using this method for our final comparison between numerical 
simulations and the theoretical prediction of the Langevin model. 
The result of this comparison for $\mu = -\delta\zeta_6$ is found 
in Table \ref{tab:best} for the small and the large cascade. In both
cases, numerics of the REWA cascade and theory agrees within error bars,
the size of the exponent corrections differing by more than a factor
of three between the small and the large cascade. This underscores 

\begin{table}[bt]
  \begin{center}
    \leavevmode
    \begin{tabular}[t]{lr@{${}={}$}l
        *{1}{d@{${}\cdot{}$}l@{${}\pm{}$}d@{${}\cdot{}$}l}}
      &\multicolumn{2}{c}{ }&\multicolumn{4}{c}{$\mu=-\delta\zeta_{6}$}
      \\ \tableline
      REWA:
      &$N$&$26$&1.3&$10^{-2}$&1&$10^{-3}$
      \\  
      Langevin:
      &$R/D^{1/2}$&$1.729\cdot10^{-1}$&1.3&\multicolumn{3}{l}{$10^{-2}$}
      \\ \tableline
      REWA:
      &$N$&$74$&5.3&$10^{-3}$&1&$10^{-3}$
      \\  
      Langevin:
      &$R/D^{1/2}$&$9.968\cdot10^{-2}$&4.2&\multicolumn{3}{l}{$10^{-3}$}
    \end{tabular}
  \end{center}
  \caption{ Comparison of our best numerical estimate of 
    of the intermittency correction $\mu = -\delta\zeta_{6}$ for
    the REWA cascades with the theoretical prediction.
    In both cases, with the predicted exponents differing by a factor
    of three, agreement is within error bars.
            }
  \label{tab:best}
\end{table}

\begin{figure}[H]
  \begin{center}
    \leavevmode
    \epsfsize=0.4 \textwidth
    \epsffile{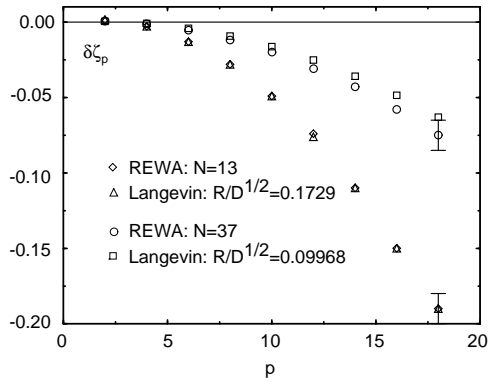}
    \caption{Intermittency corrections for the two REWA cascades 
      in comparison with the  prediction of theory.
      The error bars of the numerical values are based on a
      weighted least square fit, and are given exemplary for the 
      highest moment.
        }
    \label{fig:intermittdev}
  \end{center}
\end{figure}

the consistency of our results and demonstrates that the Langevin 
model captures all the essential physics responsible for the 
build-up of fluctuations in a local cascade. The same message is contained 
in Fig. \ref{fig:intermittdev}, which summarizes the exponent corrections
for both models. A better understanding of why a model, which has the 
energy as its only mode works so well, is supplied by a study of the 
{\it temporal} correlations. At the same time it gives insight into
the origin of intermittency itself, at least in models with  
local coupling.


\section{Temporal correlations}
\label{sec:timecorr}

In the previous sections we have looked only at equal time 
correlations. Although they contain information about the 
fluctuations of the cascade, their information about the dynamics
responsible for these fluctuations is very indirect. So ultimately
one has to look at temporal correlations as well to understand the 
dynamical origins of intermittency. In the case of the REWA cascade,
one can look at the fluctuations of the individual Fourier modes,
\begin{equation}
  \label{ucorr}
   C^{(u)}({\bf k},t)=\left<{\bf u}^{*}({\bf k},0)\cdot{\bf
         u}({\bf k},t)\right> ,
\end{equation}
just as one does in studies of the full Navier-Stokes equation. 
We also expect the temporal correlations of the total energy 
within a cascade step to be of particular significance, 
\begin{equation}
  \label{ecorr}
   C_{\ell}^{(E)}(t) = 
     \left<E_{\ell}(0)E_{\ell}(t)\right>-\left<E_{\ell}\right>^{2},
\end{equation}
since the conservation properties of the energy are responsible for 
maintaining a turbulent state. 

\begin{figure}[H]
  \begin{center}
    \leavevmode
    \epsfsize=0.4 \textwidth
    \epsffile{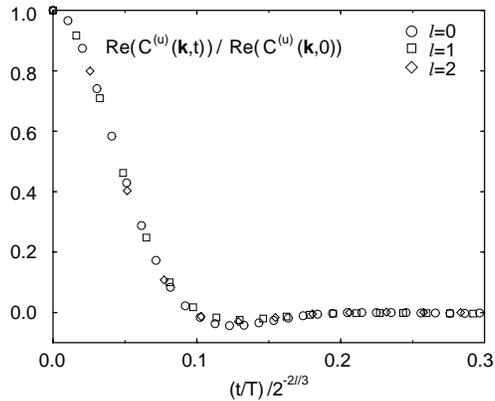}
    \caption{The real part of the autocorrelation function
      $C^{(u)}({\bf k},t)=\left<{\bf u}^{*}({\bf k},0)\cdot{\bf u}({\bf
        k},t)\right>$ normalized to its equal time value $C^{(u)}({\bf k},0)$
      for the mode ${\bf k}=2^{\ell}(1,1,1)$ of the small cascade.
      Shown are the three top levels. 
      The correlation function decays
      very quickly, reflecting the chaotic behavior of the velocity field.
       The time,
       given in units of $T = \left<E_{0}\right>/\epsilon$, is rescaled 
       according to the classical prediction $\tau^{(u)}_{\ell} 
       \sim 2^{-2\ell/3}$ for the decay time. 
                  }             
    \label{fig:modecorrelation}
  \end{center}
\end{figure}

The standard guess for the scale dependence of correlation times is 
based on the Kolmogorov picture. Namely, assuming that the typical
correlation time is a local quantity, the only combination of the 
length scale $2^{-\ell} L$ and the mean energy transfer $\epsilon$ 
having dimensions of time leads to 
\begin{equation}
  \label{classtime}
  \tau_{\ell}\sim \left[(2^{-\ell} L)^2/\epsilon\right]^{1/3}\ .
\end{equation}
We test this idea by computing the temporal correlation (\ref{ucorr})
of a particular mode of the REWA cascade, for three different shells,
as plotted in Fig. \ref{fig:modecorrelation}.
The abscissa is rescaled according to (\ref{classtime}), which makes 
all three correlation functions fall atop of each other. So it 
seems that if corrections to (\ref{classtime}) exist, they are so small 
that they need more sophisticated methods to be revealed. 

\begin{figure}[H]
  \begin{center}
    \leavevmode
    \epsfsize=0.4 \textwidth
    \epsffile{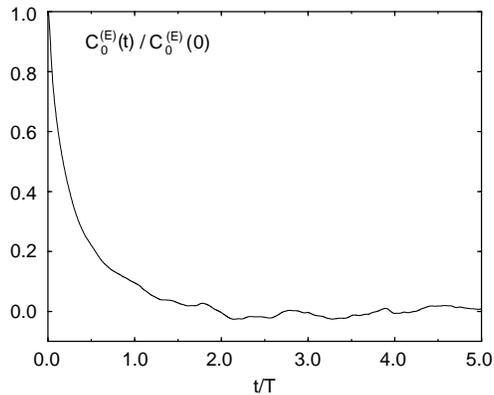}
    \caption{The autocorrelation function $C^{(E)}_{0}(t)$ of the energy on
      the top level, normalized to its equal time value
      $C^{(E)}_{0}(0)$ for the small cascade. The energy
      de-correlates by more than a factor of 10 more slowly than a 
      single velocity mode. 
       The time is given in units of
      $T = \left<E_{0}\right>/\epsilon$.}
    \label{fig:energycorrelation}
  \end{center}
\end{figure}

Next, we look at the temporal correlations of the energy on the 
highest level, plotted in Fig. \ref{fig:energycorrelation}. 
Apart from differences in shape, the remarkable feature is that 
the typical decay time is more than 10 times as long as that of 
individual Fourier modes. The temporal correlation is apparently 
dominated by the long-time fluctuations of the energy already noted 
in Fig. \ref{fig:energy}. The rapid fluctuations representative of the
individual Fourier modes appear to be completely uncorrelated. 
The physical reason is that many different modes within a shell 
contribute to these rapid fluctuations, which are de-correlated 
through many random ``collisions''. Just like in a gas of particles, 
the complicated interaction between many modes tends to randomize 
the individual motions. 

\begin{figure}[H]
  \begin{center}
    \leavevmode
    \epsfsize=0.4 \textwidth
    \epsffile{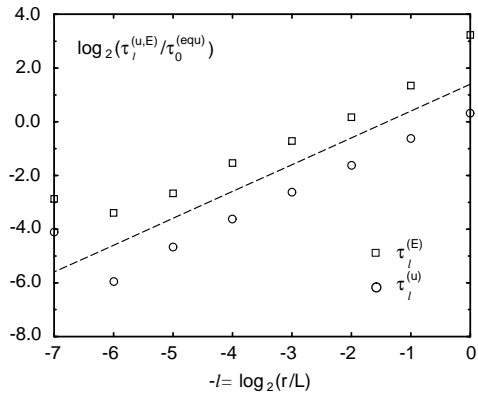}
    \caption{Correlations times $\tau^{(u,E)}_{\ell}$ 
    for the small cascade in equilibrium, as a function of the 
    level number. The dashed line corresponds to the theoretical
    prediction of (\protect{\ref{equtime}}). The time scale of
    the energy is longer than that of an individual mode 
    by a factor of 4 on all levels.
       }
    \label{fig:equtime}
  \end{center}
\end{figure}

If the small-scale fluctuations were a true random walk, the energy 
would slowly diffuse to take arbitrary values. But eventually the 
tendency of the dynamics to restore equilibrium will drive the energy 
back to its mean value. This requires a coherent motion of many 
individual Fourier modes, whose correlations need some time to build up. 
Thus the long time scale in the motion of the energy. So far the same
argument would apply for a cascade in equilibrium, in the absence 
of driving. This is shown in Fig. \ref{fig:equtime} for a cascade of
8 shells, 
which all perform fluctuations whose static distribution
was shown in Fig. \ref{fig:equilib}. Since the energies of all shells 
fluctuate around some common value $E_{av}$, and the local length 
scale is $2^{-\ell} L$, for dimensional reasons the time scale must be 
\begin{equation}
  \label{equtime}
  \tau_{\ell}^{(equ)} \sim E_{av}^{-1/2} \;  2^{-\ell} L \; .
\end{equation}
The correlation times shown in Fig. \ref{fig:equtime} for both the 
energy and one of the velocity modes is calculated according to
\begin{equation}
  \label{decorrelation}
     \tau_{\ell}^{(E,u)} = 
     \frac{\int_{0}^{\infty} C_{\ell}^{(E,u)}(t)\, dt}
     {C_{\ell}^{(E,u)}(0)} \quad,
\end{equation}
and are found to corroborate the scaling law (\ref{equtime}). 
As in the case of the top level of a Kolmogorov cascade, the correlation
time of the energy is slower compared to an
individual Fourier mode. 

\begin{figure}[H]
  \begin{center}
    \leavevmode
    \epsfsize=0.4 \textwidth
    \epsffile{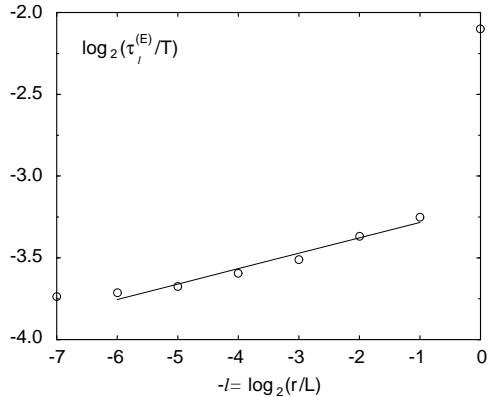}
    \caption{Scaling of the correlation times 
    $\tau_{\ell}^{(E)}$
       of the energy correlation function $C^{(E)}_{\ell}(t)$ 
      for the small cascade.
      The fit corresponds to $\tau^{(E)}_{\ell} \sim 2^{-0.094\ell}$,
      which is a very small decrease compared to the classical 
      prediction of $\tau^{(E)}_{\ell} \sim 2^{-2\ell/3}$.
          }
    \label{fig:tscale}
  \end{center}
\end{figure}

This changes fundamentally when looking at lower shells of a 
non-equilibrium cascade. For a turbulent cascade, one expects to 
recover the scaling (\ref{classtime}), since the correlation of the
energies are defined in terms of local quantities only. But remarkably,
the scaling law (\ref{classtime}) does not even approximately describe
the scaling of $\tau^{(E)}_{\ell}$, which rather follows the 
power law
\begin{equation}
  \label{strange}
    \tau^{(E)}_{\ell} \sim 2^{-\alpha\ell} \;, \alpha = 0.094,
\end{equation}
as seen in Fig. \ref{fig:tscale}.

This means the time scale of the motion of the energy hardly gets
shorter at small scales. The reason lies within the non-equilibrium
properties of the cascade: Once a large fluctuation of the energy 
has built up on the highest level, it can only relax by being 
transported to a lower level. Thus the same long time dependence 
is imprinted on the lower level, which again can only be transferred to 
the next level. Thus on a given level, {\it all} time scales of the 
levels lying above appear, and apart form the single rescaling 
(\ref{strange}) the {\it shape} of the correlation function changes as well. 
This corresponds directly to the origin of intermittent fluctuations 
itself: fluctuations amplify because lower shells are driven by 
slowly varying energy input of the higher shells. Consequently
fluctuations ``ride'' atop of the long-scale fluctuations and amplify.
By looking at correlations between individual 
Fourier modes like (\ref{ucorr}) none of these long-range 
correlations are revealed. 
Correlations between the coherent structures of a turbulent velocity
field will always be masked by the incoherent fluctuations of the 
individual modes.

It is interesting to note that in a numerical simulation of isotropic 
turbulence Yeung and Pope \cite{yeung89} have found very strong deviation
from classical scaling as well. They looked at the Reynolds number 
dependence of the acceleration variance, which is a Lagrangian 
quantity. Like in our case, including fluctuations of the energy 
transfer according to Kolomogorov's refined similarity hypothesis 
\cite{kolmogorov62} cannot account for the corrections found. 
However, a meaningful definition of a Lagrangian quantity within our 
local approximation is difficult, since convection cannot 
be described properly. Therefore we do not know how to relate 
our findings with the results of Ref. \cite{yeung89} directly. 

The second consequence of the extremely slow de-correlation of the 
energy is
that the ratio of time scales of the energy and that of individual 
modes rapidly becomes larger on smaller scales. In the Langevin 
description the white noise term represents the irregular motion of
individual Fourier modes, while our main interest lies with the 
large-scale fluctuations of the energy. In the scaling limit we are 
interested in the disparity between these scales becomes infinitely
large. Thus a white-noise description should become better and better
in the relevant limit.



\section{Discussion}
\label{sec:dissc}

We have investigated mode-reduced approximations of the 
Navier-Stokes equation and found them to have anomalous scaling exponents 
in the inertial range. Intermittent fluctuations come about through 
rare excursions of the energy from its mean value, which originate 
from the top levels of the cascade. They are thus well described 
by a white-noise Langevin process, whose random forcing represents the 
motion of individual Fourier modes. The analytical solution shows
\cite{eggers94} that such a Langevin cascade necessarily exhibits corrections
to the classical scaling exponents. It is thus hard to see how
{\it any} cascade with local coupling could avoid intermittency 
corrections in the inertial range, since for a chaotic motion there
will always be fluctuations, and thus $D/R^2 \neq 0$.
In fact, as pointed out by 
Kraichnan \cite{kraichnan74}, the only mechanism by which such 
fluctuations could be {\it avoided} is by sufficiently strong mixing
in space, which in $k$-space would be represented by non-local 
interactions. The fact that our simple Langevin model gives an
excellent description of a complicated network of Fourier modes
is also highlighted by the excellent agreement in an equilibrium 
state. We have recently extended the comparison between the 
Langevin and the REWA model to the formation of singularities 
in the absence of viscosity \cite{uhlig96}. In this case, 
which represents a state even farther from equilibrium than 
a Kolmogorov cascade, the two models still compare quantitatively.
This allows for an analytical description of the singularities 
of the Euler equation \cite{majda90} in the approximation of local 
coupling.

The second central point of our paper is to demonstrate that 
intermittency exponents can be calculated for a complex cascade with 
nonlinear, chaotic dynamics by analytical means. The only piece of
information one needs about the turbulent flow concerns the {\it local}
energy transfer. With this information, the global 
non-equilibrium properties determine the 
intermittency exponents, which is taken care of by the analytical 
calculation. In a very interesting paper, Olla \cite{olla95} has 
recently carried out the same program directly 
from the Navier-Stokes equation. He assumes that the broadening 
of the distribution of the velocity field from a scale $r$ to $r/2$
is adequately described in a Gaussian approximation. Thus he is 
able to determine the analogue of the noise strength $R/D^{1/2}$
from perturbation theory. However, a major problem of Olla's 
work is that he also does not deal adequately with the spatial structure 
of turbulence, but rather maps the problem on the same linear cascade
structure we are considering here. 

The most pressing problem remaining with the present mode
representations is that the intermittency effects are quite small.
Thus some important physical features are still missing. The most 
likely reason is the fact that a given eddy only feeds a single 
substructure, instead of branching out to several smaller structures
overlapping with it in physical space. It has been shown 
\cite{eggers91b,eggers92} that such a tree structure may strongly 
enhance intermittent 
fluctuations, owing to competition between different eddies on the
same level. The procedures of the present paper can be directly carried over
to this physical situation, as analytical solutions of the Langevin
model with tree structure are available as well \cite{uhlig94}. 
Unfortunately, there is no technique available which would link the 
Navier-Stokes equation to a cascade coupled locally in real 
and in $k$-space in a rational way. Since the value of intermittency 
exponents depends significantly on the properties of the local couplings 
\cite{grossmann92}, no quantitative prediction for real three-dimensional 
turbulence is possible using the more general tree structure. 

An interesting fact revealed by our analysis is that the value 
of the correction exponents depends considerably on the mode selection. 
There is a tendency, observed earlier \cite{grossmann94a}, for the 
size of fluctuations to decrease with the number of modes per shell. 
In \cite{grossmann94a}, mode systems with up to 86 modes per shell have
been considered, which also allow for more distant interactions between 
shells. Although it is interesting to look at the significance of 
more distant interactions, it does not remedy the fundamental 
problem of the present approximation, which is that the number of 
modes does not proliferate on small scales. In terms of interactions, 
there is still no competition between modes of the same scale. Thus 
no conclusions about the true intermittency exponents can be drawn
from including more modes, as long as the spatial structure is 
not respected. In fact for a tree structure the number of localized 
boxes would increase like $2^{3\ell}$, each with a constant number 
of modes $N$. Comparing that with the total number of Fourier modes
$(4\pi/3)(2^3 - 1)2^{3\ell}$ within a shell for three-dimensional 
turbulence, one obtains the estimate 
$N \approx 28$, which is close to the number $N  = 26$ of our small
cascade \cite{Siggia77}. Thus the small cascade already represents a 
physically reasonable choice for the number of locally interacting 
Fourier modes.

Finally we mention the so-called GOY model \cite{jensen91,benzi95},
which contains only one complex mode per shell,
and which has attracted considerable attention recently. Although it has 
no spatial structure, the intermittency exponents are comparable 
to those of real turbulence, which seems to contradict our 
above observations. We believe that the mechanisms at work are 
somewhat different here, since there exist pulses, which run down 
the cascade. These global coherent motions are much more effective
in transporting fluctuations to smaller scales than in REWA cascades, 
where such structures are destroyed by intra-shell mixing. A separation 
of time scales between the total energy and the individual modes 
of a shell evidently does not exist. Another fact which points to
global structures is the dependence of intermittency exponents 
on the form of viscous dissipation \cite{benzi95}. Still,
this simple model remains interesting in particular in view of 
the analytical theory which has been developed for its inertial 
range properties \cite{benzi93c}, and which has a number of similarities
with the one developed for the Langevin model \cite{eggers92,eggers94}.

To conclude, we have investigated in detail how local coupling leads 
to intermittent fluctuations. In spite of many other interesting proposals, 
Landau's and Kolmogorov's original arguments remain among the 
most powerful for their understanding.


\section*{Acknowledgments}
This work is supported by the Sonderforschungsbereich 237 (Unordnung und
grosse Fluktuationen).


\end{document}